\documentclass[]{mn2e}

\usepackage[dvips]{graphicx,color}
\usepackage{times}

\usepackage{amsmath}
\usepackage{pifont}
\usepackage{lscape}

\begin{document}

\title[NGC\,300 X--1 is a WR/black-hole binary]
{NGC\,300 X--1 is a Wolf-Rayet/Black-Hole binary\thanks{Based on observations
made with ESO telescopes at the Paranal Observatory under programme
ID 384.D-0093(A)}}
\author[P. A. Crowther et al.]{P. A. Crowther$^{1}$\thanks{Paul.Crowther@shef.ac.uk},  
R. Barnard$^{2}$, S. Carpano$^{3}$, J. S. Clark$^{2}$, V. S. Dhillon$^{1}$, 
A. M. T. Pollock$^{3}$
\vspace{3mm} \\ 
$^{1}$Department of Physics and Astronomy, University of Sheffield, 
Sheffield S3 7RH, UK\\
$^{2}$ Department of Physics and Astronomy, The Open 
University, Walton Hall, Milton Keynes, MK7 6AA, UK\\
$^{3}$ XMM-Newton Science Operations Center, ESAC, 28080, Madrid, Spain}
\date{\today}

\pagerange{\pageref{firstpage}--\pageref{lastpage}} \pubyear{2009}

\maketitle

\label{firstpage}

\begin{abstract} We present VLT/FORS2 time-series spectroscopy of the 
Wolf-Rayet star \#41 in the Sculptor group galaxy NGC\,300. We confirm a 
physical association with NGC\,300 X--1, since radial velocity variations 
of the He\,{\sc ii} $\lambda$4686 line indicate an orbital period 
of 32.3 $\pm$ 0.2 hr which agrees at the 2$\sigma$ level with
the X-ray period from Carpano et al. We measure a 
radial velocity semi-amplitude of 267$\pm$8 km\,s$^{-1}$,  from which a 
mass function of 2.6 $\pm$ 0.3 $M_{\odot}$ is obtained. A  revised 
spectroscopic mass for the WN-type companion of 26$^{+7}_{-5}$ 
$M_{\odot}$ yields a black hole mass of 20 $\pm$ 4 $M_{\odot}$ for 
a preferred inclination of $60-75^{\circ}$. If the WR  star 
provides 
half of the measured visual continuum flux, a reduced WR (black hole) mass 
of 15$^{+4}_{-2.5}\,M_{\odot}$ (14.5$^{+3}_{-2.5}\,M_{\odot}$) would be 
inferred. As such, \#41/NGC\,300 X--1 represents only  the second 
extragalactic Wolf-Rayet plus black-hole binary system, after 
IC\,10 X--1. In addition, the compact object responsible for NGC\,300 X--1 
is the second highest stellar-mass black hole known to date, exceeded only 
by IC\,10 X--1. 
\end{abstract}


\section{Introduction}

High mass X-ray binaries (HMXB) typically comprise OB stars plus either a 
neutron star or black hole, in which high X-ray luminosities ($\sim$10$^{38}$ 
erg\,s$^{-1}$) arise from accretion disks around the compact object. 
Accretion disks are fed through a combination of Roche lobe overflow and 
stellar winds from the early-type companion. Very few known HMXB are known 
to host black-holes, Cyg X--1 in the Milky Way, X--1 and X--3 in the Large 
Magellanic Cloud and X--7 in the Local Group galaxy M\,33. If a merger
is avoided, the OB companion 
will potentially evolve through to a Wolf-Rayet phase, producing a system 
comprising a  helium star plus a black hole or neutron star (Tutukov \& 
Yungelson 1973). Indeed, van den Heuvel \& de Loore (1973) proposed the 
Galactic HMXB Cyg X--3 as a helium-star plus compact object system, 
which was
observationally confirmed by van Kerkwijk et al. (1992). Such systems are 
believed to be very rare, with as few as $\sim$100 helium star plus black 
hole pairs in the Galaxy (Ergma \& Yungelson 1998).

\begin{table}
\begin{center}
\caption{Log of VLT/FORS2 spectroscopic observations of \#41 in NGC\,300.
UT dates and MJD's refer to the start of the 1,535~s exposures. We include 
individual radial velocities, $v_{\rm r}$, as measured from Gaussian 
fits to He\,{\sc ii} $\lambda$4686. 
Phases adopt a period of 32.3 hr, where phase 0 refers to MJD 55118.97559$\pm$0.01554).}
\begin{small}
\begin{tabular}{c@{\hspace{3mm}}r@{\hspace{1.5mm}}c@{\hspace{1.5mm}}c
@{\hspace{2mm}}c@{\hspace{0.5mm}}c@{\hspace{0.5mm}}c@{\hspace{2mm}}c}
\hline 
UT Date & MJD & DIMM & sec z & \multicolumn{3}{c}{$v_{\rm r}$ $\lambda$4686} & 
Phase\\
        &    --55119       & $''$ &       & \multicolumn{3}{c}{km\,s$^{-1}$} & \\
\hline
01:06 15 Oct 09 &0.046180 &1.2&1.29&  347.0 & $\pm$ &  17.7 & 0.052\\
05:28 15 Oct 09&0.227922 &2.3&1.08&  364.6 & $\pm$ &  25.5 & 0.187\\
01:40 16 Oct 09&1.069495 &1.1&1.18 &  --64.2 & $\pm$ &  16.9 & 0.812\\
05:58 16 Oct 09&1.249073 &1.7&1.14 &  136.0 & $\pm$ &  25.0 & 0.945\\
04:07 19 Oct 09&4.171590 &0.7&1.03&  369.0 & $\pm$ &  14.3 & 0.114\\
08:24 19 Oct 09&4.350382   &0.8&1.98& 456.6 & $\pm$ &  20.1 & 0.247\\
02:31 20 Oct 09&5.104896 &0.9&1.06 & --68.3 & $\pm$ &  12.1 & 0.807\\
06:41 20 Oct 09&5.278856 &0.7&1.31 &  109.4 & $\pm$ &  17.3 & 0.936\\
00:14 25 Oct 09&10.010318 &1.2&1.34 & 325.3 & $\pm$ &  25.7 & 0.447\\
06:28 25 Oct 09&10.269520 &0.7&1.33&  10.4 & $\pm$ & 14.3 & 0.639\\
\hline
\end{tabular} 
\label{log}
\end{small}
\end{center}
\end{table}

To date, the only confirmed Wolf-Rayet plus black hole binary system is 
IC\,10 X--1 in the dwarf irregular galaxy IC\,10. This system has a period 
of 34.9 hr, hosts a WN-type Wolf-Rayet star [MAC 92] 17A (Crowther et al. 
2003) and an unseen companion which is currently the record holder amongst 
stellar mass black holes, exceeding 23.1$\pm$2.1 M$_{\odot}$ (Prestwich et 
al. 2007; Silverman \& Filippenko 2008). 

\begin{figure}
\begin{center}
\includegraphics[width=0.75\columnwidth,clip]{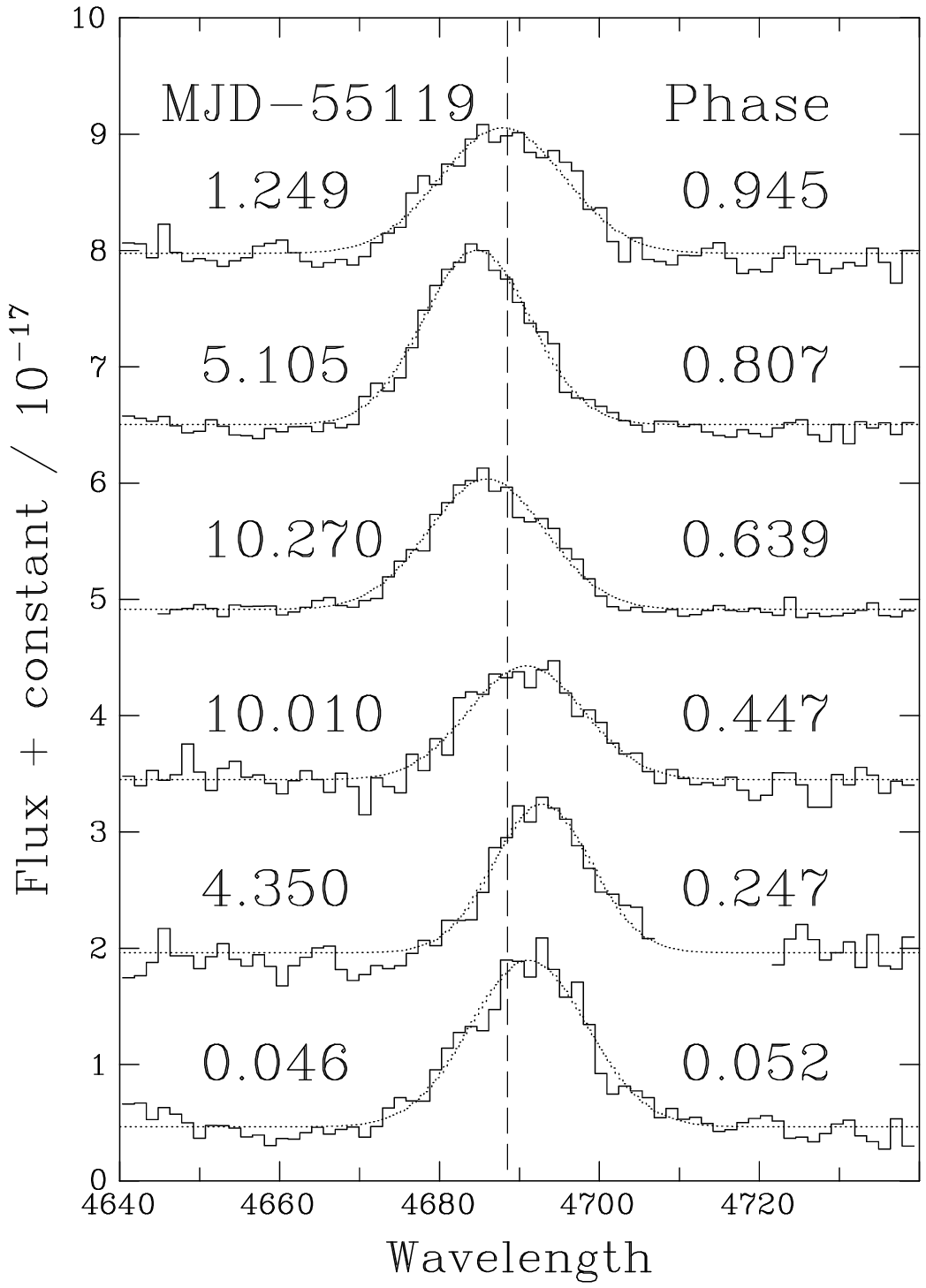}
\caption{Representative VLT/FORS2 spectroscopy of \#41 revealing
radial velocity variations in He\,{\sc ii} $\lambda$4686 (gaussian
fits are shown as dotted lines), successively offset by 1.5$\times
10^{-17}$ erg\,s$^{-1}$\,cm$^{-2}$\,\AA$^{-1}$ for clarity.}
\label{4686}
\end{center}
\end{figure}

A second candidate extragalactic black hole plus Wolf-Rayet system is 
NGC\,300 X--1 (Carpano et al. 2007a), in the southern Sculptor group 
spiral galaxy NGC\,300 which lies at a distance of 1.88~Mpc (Gieren et al. 
2005). This source is spatially coincident with \#41 from the Wolf-Rayet 
catalogue of Schild et al. (2003) which was confirmed as a WN-type 
Wolf-Rayet star by Crowther et al. (2007). Both systems exhibit similar 
X-ray properties (Carpano et al. 2007b; Barnard, Clark \& Kolb 2008).
However, no evidence exists to date of a physical 
link between NGC\,300 X--1 and the Wolf-Rayet star \#41. This is the 
purpose of the present Letter.

New Very Large Telescope (VLT) optical spectroscopic time-series of \#41 
are discussed in \S\,\ref{section2} in which variations are revealed in 
the 
He\,{\sc ii} $\lambda$4686 emission line.  \S\,\ref{section3} compares the 
inferred orbital period with X-ray lightcurves and derives a 
semi-amplitude for the Wolf-Rayet star, from which a mass function is 
obtained. \S\,\ref{section4} provides a revised mass for the Wolf-Rayet 
star, placing strict limits upon the mass of the compact companion. We
conclude with  a brief discussion in \S\,\ref{section5}.

\section{Observations} \label{section2}

Here we present new VLT optical spectroscopy of \#41 obtained with the 
Focal Reducer/Low Dispersion Spectrograph \#2 (FORS2) in multi-object
spectroscopy (MOS) mode from 14--24 October 2009. Two 1,535~s exposures
were obtained on each of five non-consecutive nights using the 600B grism,
centred at 465\,nm. A log of our observations is presented in Table~\ref{log}
including DIMM seeing measurements, which indicated typical seeing of
0.7--1.2 arcsec, except at two epochs for which the seeing exceeded
1.5 arcsec.

MOS allows 19 sources to be simultaneously observed. We included 7 H\,{\sc 
ii} regions in NGC\,300 plus 12 WR candidates from Schild et al. (2003). Of 
these, 3 sources have previously been spectroscopically confirmed as WR 
stars, namely \#9, \#41 and source 12 from Bresolin et al. (2009). Other 
sources will be discussed elsewhere.

After bias subtraction and flat field correction, a standard extraction 
was performed using {\sc iraf}, with wavelength and flux calibration 
carried out using {\sc figaro}.  1.0 arcsec slits provided a spectral 
resolution of 4.2\,\AA\ -- as measured from comparison arc lines -- with a 
wavelength coverage of 3300--5800\AA\ for \#41. Faint H$\beta$ and 
[O\,{\sc iii}] $\lambda\lambda$4959, 5007 nebular lines were detected in the \#41 
slit. \#41 does not appear to be responsible for these features since
the intensity of the nebula is weak and spatially close to \#41, and
has a lower systemic velocity of $\sim$95 km\,s$^{-1}$ (versus  202
km\,s$^{-1}$ for \#41).

Spectrophotometric standard stars were observed with FORS2 in long slit 
mode on the nights on 15 Oct (HD\,49798), 18--19 Oct (LTT\,2415) and 24 
Oct (LTT 7987), providing a wavelength coverage of 3300-6230\AA. An 
absolute flux calibration was achieved for \#41 using B = 22.71 mag from 
Carpano (2006).

\begin{figure}
\begin{center}
\includegraphics[width=0.6\columnwidth,angle=270,clip]{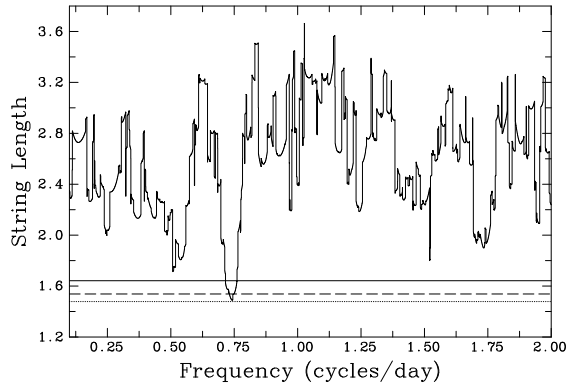}
\caption{String-length periodogram of the VLT/FORS2 radial velocities. The
minimum string length indicates the best period, at a frequency of 0.7422 cycles/day.
The solid, dashed and dotted lines represent the 68.3\%, 95.4\% and 99.7\%
confidence levels, respectively (see text for details).}
\label{string}
\end{center}
\end{figure}

\section{Orbital period}\label{section3}

Several VLT/FORS2 spectroscopic
observations in the vicinity of He\,{\sc ii} $\lambda$4686 
are presented in Fig.~\ref{4686}, revealing large radial velocity variations.
Gaussian profiles are fit to individual $\lambda$4686 profiles, with
individual centroids listed in Table~\ref{log}.


In view of the sparsely sampled datasets, we have employed the
string-length approach of Dworetsky (1983). The data are folded on a
set of trial frequencies and the total length of ``string'' required
to join the observations in phase order is calculated.  The smallest
string length found from the search is assumed to correspond to the
correct period. The resulting periodogram is shown in Fig.~\ref{string}. 
The
deepest trough corresponds to a period of $32.3\pm0.2$\,hr. The error
on this period was computed by constructing 10\,000 synthesized
datasets and measuring the standard deviation of the positions of the
deepest troughs in the resulting periodograms. The synthesized
datasets were obtained by ``jiggling'' each data point about its
observed value by an amount given by its error bar multiplied by a
number output by a Gaussian random-number generator with zero mean and
unit variance.

It can be seen that our derived period is in agreement at the 2$\sigma$
level with the Swift X-ray period of $32.8\pm0.2$\,hr (1$\sigma$;
Carpano et al. 2007b), which gives us confidence that we have
identified the correct value. Moreover, the minimum string length of
our period (1.49) compares favourably with the string length of a
perfect sinusoid (1.46) and the string length of a sinusoid with noise
consistent with the error bars on the observed data added to it
(1.48). 

To further test the significance of our derived period, we
used a randomization technique (Fisher 1935). The radial velocities
were randomly reassigned to the times of observation, thereby
preserving the data sampling and the mean and standard deviation of
the original dataset. A set of 10\,000 randomised datasets were
constructed in this way and then subjected to the same string-length
periodogram analysis.  By constructing a cumulative distribution
function of the resulting minimum string lengths, we are able to place
confidence limits on the significance of a given string length, as
shown by the horizontal lines in Fig.~\ref{string}. We are able to reject 
the hypothesis that our preferred period is due to noise at $99.43$\%
confidence and eliminate the next highest troughs as most likely due
to noise. Note that the troughs around 0.25 and 1.75 cycles/day are
due to the one-cycle-per-day alias. As a check on the string length
method, we also computed a Lomb-Scargle periodogram (Press \& Rybicki
1989) and obtained consistent results.

\begin{figure}
\begin{center}
\includegraphics[width=0.6\columnwidth,angle=270,clip]{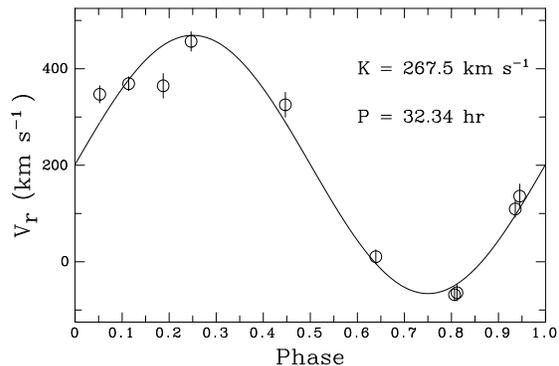}
\caption{Radial velocity variations of $\lambda$4686 He\,{\sc ii}
phased to 32.3 hr,  
from which a systemic velocity of  $v_{\rm r}$ = 202 $\pm$ 7 
km\,s$^{-1}$ and semi-amplitude of  $K_{2}$ = 267.5 $\pm$ 7.7 km\,s$^{-1}$ is obtained.}
\label{sine}
\end{center}
\end{figure}

Adopting our optical period, we present the phased radial velocity 
measurements in Fig.~\ref{sine}, with a 
semi-amplitude of $K_{2}$ = 267.5$\pm$7.7 km\,s$^{-1}$. 
In the Figure, phase 0 corresponds to 
MJD 55118.97559$\pm$0.01554.

Armed with the semi-amplitude and orbital period, we may now derive the 
system mass function, \[ f(m) = \frac{P K_{2}^{3}}{2 \pi G} = \frac{M_{1} 
\sin^{3} i}{(1+q)^2} \] where $M_{1}$ is the compact object mass, and $q = 
M_{2}/M_{1}$. The derived mass function is $f(m) =  2.63\pm0.33$
$M_{\odot}$, and would 
correspond to the compact object mass in the case of a system viewed at an 
inclination of 90$^{\circ}$ whose companion mass is negligible. In the 
case of a massive companion star with $q \sim 1$, the compact object would 
have a minimum mass of 4 $f(m)$. As such, the  compact object in NGC\,300 
X--1 is indeed a black hole, such that this 
system represents only the second confirmed WR plus black hole binary.

\section{Wolf-Rayet properties}\label{section4}

\begin{figure}
\includegraphics[width=0.7\columnwidth,angle=270,clip]{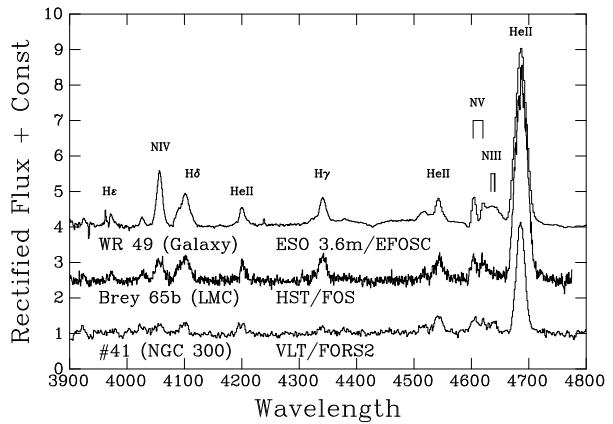}
\caption{Comparison between rectified, phase corrected VLT/FORS2 spectrum of \#41 
with Galactic (WR 49) and LMC (Brey 65b = NGC 2044 West 5C) 
weak-lined WN5 stars, respectively from Hamann et al. (1995) and Walborn et al. (1999).}
\label{ngc300x1_rect}
\end{figure}

We present our new, combined (phase-corrected), rectified 
VLT/FORS2 spectrum of \#41 in Fig.~\ref{ngc300x1_rect}.
This high quality spectrum confirms a weak-lined WN5 subtype, 
previously inferred by Crowther et al. (2007) from lower 
resolution, lower S/N spectroscopy obtained with VLT/FORS2 using 
the 300V  grism in January 2007.  Overall, the visual spectrum of \#41 is 
similar to other weak-lined WN5 stars, namely WR 49 in the Milky Way and Brey 65b
(= NGC 2044 West 5C) in the LMC, taken from  
Hamann et al. (1995) and Walborn et al. (1999),  respectively. The 
He\,{\sc ii} $\lambda$4686 equivalent width ($W_{\lambda} \sim$56\AA), and 
line width (FWHM$\sim$17\AA) in \#41  are somewhat lower than LMC and 
Milky Way counterparts, $W_{\lambda}$=110--140\AA\ and FWHM=22--24\AA.

In order to reassess the mass of \#41 we have calculated a synthetic model 
using the Hillier \& Miller (1998) line-blanketed, non-LTE model 
atmosphere code. With respect to Crowther et al. (2007), somewhat more 
sophisticated atomic models are considered, namely H, He, C, N, O, 
Ne,  Si, P, S, Ar, Fe and Ni. Elemental abundances are set to 40\% of the 
solar  value (Urbaneja et al. 2005), with the exception of H and CNO 
elements. Clumping is accounted for, albeit in an approximate manner, with a 
(maximum)  volume filling factor of 10\%, such  that the derived mass-loss 
rate is  three times smaller than the value that would have been obtained
by assuming an homogeneous wind.

In view of the weak He\,{\sc i} line spectrum in \#41, we have based our 
analysis upon He\,{\sc ii} ($\lambda$4686, $\lambda$5411) and N\,{\sc 
iv-v} ($\lambda\lambda$4603--20, $\lambda$4058, $\lambda$7103--7129) line 
diagnostics. Overall good agreement is found, which is remarkable in view
of the close proximity of the black hole to \#41. The only significant
discrepancies are  that N\,{\sc iii}  $\lambda$4634--41 is not 
reproduced in the synthetic  spectrum and excess emission is observed in 
the upper Pickering-Balmer series, the latter potentially arising from 
the accretion disk.

\begin{figure}
\includegraphics[width=0.99\columnwidth,clip]{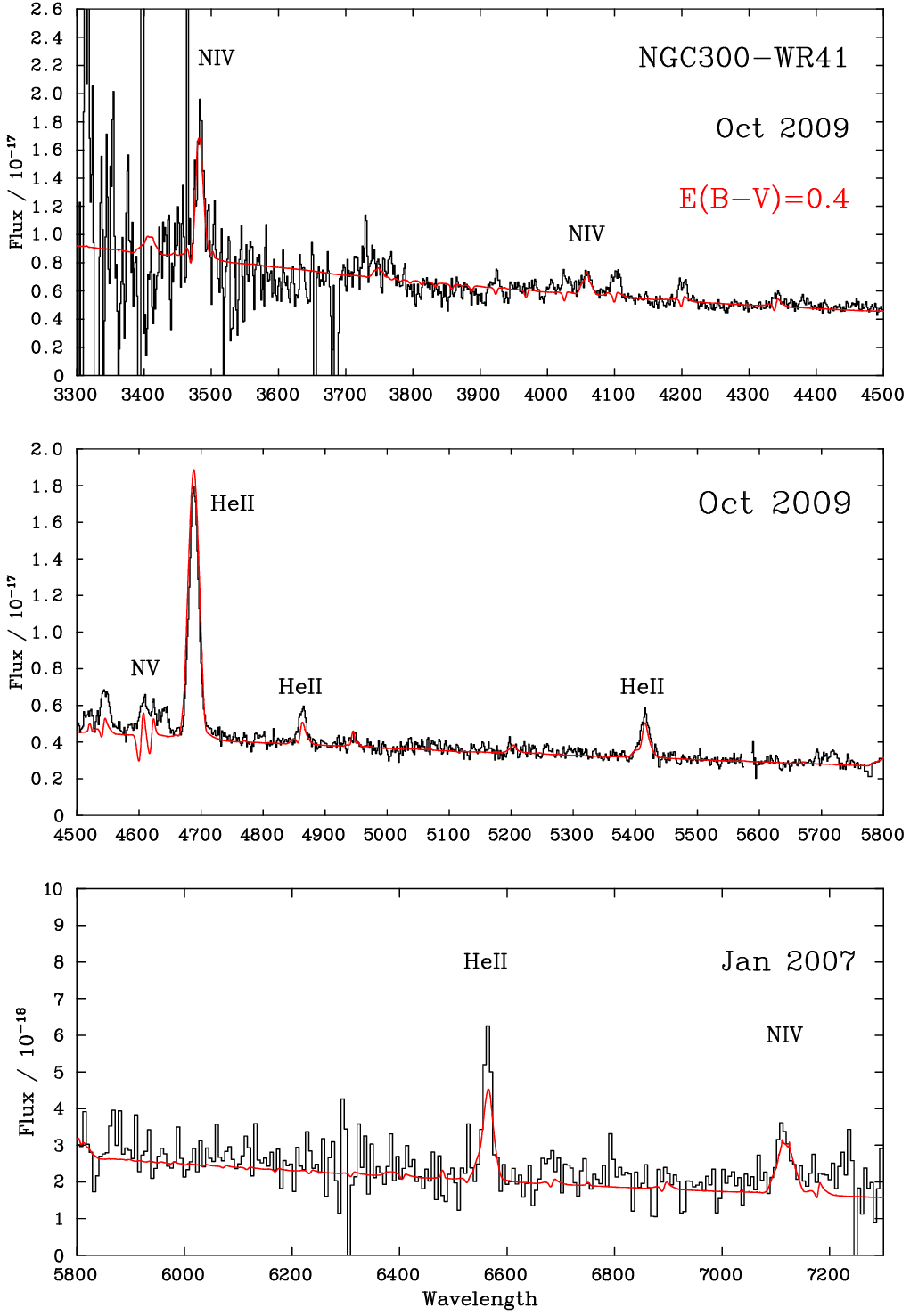}
\caption{Combined, phase-corrected VLT/FORS2 spectroscopy of 
\#41 
from Oct 2009 and Jan 2007 (black) with synthetic spectrum overlaid (red),
reddened by $E(B-V)$=0.4 mag, as described in \S\,\ref{section4}.}
\label{wr41}
\end{figure}

In Fig.~\ref{wr41} we present our new combined flux calibrated spectrum of
\#41, together with re-calibrated spectroscopy from Crowther et al. (2007)
for $\lambda >$ 5800\AA. An 
optimum fit to the spectrum of \#41 is included in the figure, and reveals the 
following stellar parameters -- $T_{\ast} \sim$ 65\,kK, $\log 
(L/L_{\odot})\sim$ 5.92, $\dot{M} \approx 5\times 10^{-6}$ M$_{\odot}$ yr$^{-1}$, 
$v_{\infty} \sim$ 1300 km\,s$^{-1}$, plus a nitrogen mass fraction of 
$\sim$0.5\%, with negligible hydrogen adopted. 
With respect to Crowther et  al. (2007), the main revision relates to a reduced 
absolute magnitude of  $M_{\rm V}$ = --5.0 mag, on the basis of a lower interstellar
reddening of $E(B-V)$ = 0.4 mag. $T_{\ast}$ should be reliable to $\pm$5\,kK,
resulting in uncertainties of $\pm$0.2 mag in bolometric corrections. Together with
$\pm$0.05 mag uncertainties in $E(B-V)$, stellar luminosities should be reliable to
$\pm$0.14 dex.

From our derived parameters, we obtain a spectroscopic WR mass of 
26$_{-5}^{+7}$ M$_{\odot}$ on the basis of the Schaerer \& Maeder (1992) 
mass-luminosity relation for hydrogen-free WR stars. The principal 
uncertainty in our inferred WR mass relates to the absolute visual 
magnitude of the WR star. Since our adopted visual magnitude is based upon 
ground-based imaging, it is possible that other continuum sources are 
included in the photometry. In this case, the WR emission line equivalent 
width would be diluted by the continuum of other nearby sources. 

Indeed, 
while  \#41 closely resembles other weak-lined WN5 stars (recall 
Fig.~\ref{ngc300x1_rect}), its He\,{\sc ii} $\lambda$4686 equivalent width 
is indeed lower by a factor of $\sim$2. This hints at a potential factor 
of two line  dilution from unresolved companions along the sight-line 
towards \#41. In this case, the WR luminosity would be reduced to $\log 
(L/L_{\odot})$ = 5.57$\pm$0.14 with the mass-loss rate unaffected,
implying a spectroscopic mass of 15$^{+4}_{-2.5} M_{\odot}$. For 
reference,
Hamann,  Gr{\"a}fener, \& Liermann (2006) estimated spectroscopic masses
of 15--19 $M_{\odot}$ for weak-lined WN5 stars in the Milky Way. In view 
of these issues,  we shall evaluate black hole 
masses using values of both 15 and 26 $M_{\odot}$ for 
the WN star.

\begin{table}
\begin{center}
\caption{Derived black hole mass, $M_{1}$, in NGC\,300 X--1 for
$i$=45, 60$^{\circ}$ or 90$^{\circ}$, for cases in 
which the WN star contributes either 50\% ($M_{2} = 15 M_{\odot}$) 
or 100\% ($M_{2} = 26 M_{\odot}$) of the visual light.
We include the separation between the components, $a$, and the 
WR radius, $R_{2}$, as a fraction of the Roche lobe radius, $r_{L}$ (Eggleton 1983).}
\begin{tabular}{c@{\hspace{2mm}}c
@{\hspace{2mm}}c@{\hspace{2mm}}c
@{\hspace{2mm}}c@{\hspace{2mm}}c}
\hline 
$M_{2}$ (WR) & $R_{2}$ (WR) & $i$ & $M_{1}$ (BH) & $a$ & $R_{2}/r_{L}$ \\
$M_{\odot}$ & $R_{\odot}$ & & $M_{\odot}$ & $R_{\odot}$ & \\
\hline
15$_{-2.5}^{+4}$   & 4.8$\pm$0.1   & 45$^{1}$   &   21.5$^{+2.5}_{-1.7}$ & 
17.0$^{+0.9}_{-0.7}$ & 0.81 \\ 
15$_{-2.5}^{+4}$   & 4.8$\pm$0.1   & 60\phantom{1}          &   
15.6$^{+1.9}_{-1.4}$ & 16.1$^{+1.0}_{-0.7}$ & 0.80  \\ 
15$_{-2.5}^{+4}$   & 4.8$\pm$0.1   & 90$^{1}$    &   12.6$^{+1.6}_{-1.3}$ 
& 15.5$^{+1.0}_{-0.8}$ & 0.79 \\ 
\\
26$_{-5}^{+7}$   & 7.2$\pm$0.1   & 45$^{1}$   &   27.8$^{+3.5}_{-2.8}$  & 
19.4$^{+1.2}_{-1.0}$ & 0.99 \\ 
26$_{-5}^{+7}$   & 7.2$\pm$0.1   & 60\phantom{1} & 20.6$^{+2.9}_{-2.1}$ & 
18.5$^{+1.3}_{-1.0}$& 0.98 \\
26$_{-5}^{+7}$   & 7.2$\pm$0.1   & 90$^{1}$      &    16.9$^{+2.4}_{-1.9}$  & 
18.0$^{+1.2}_{-1.0}$ & 0.96 \\
\hline
\end{tabular} 
\label{ngc300x1}
\end{center}
$^{1}$ The apparent glancing eclipse of the X-ray emitting accretion disk 
suggests $i = 60-75^{\circ}$.
\end{table}

\section{Discussion and Conclusions}\label{section5}

In Table~\ref{ngc300x1} we present black hole masses for inclinations of 
45, 60 and 90$^{\circ}$ for our favoured 26 $M_{\odot}$ spectroscopic 
WR mass, plus the  15  $M_{\odot}$ case, resulting from the WR star 
contributing 50\% of the visual continuum light.  We include
the separation between the components using Kepler's third law in each 
case, obtained from
\[ P = \frac{2 \pi a^{3/2}}{\left[G(M_{1} + M_{2})\right]^{1/2}},\]
corresponding to only 2.5--3.5 Wolf-Rayet radii.
%
%
The mass accretion rate required to sustain $L_{\rm X}$ = 2 $\times 
10^{38}$ erg\,s$^{-1}$ is 3.5 $\times 10^{-8}  M_{\odot}$\,yr$^{-1}$, 
if the adopted efficiency of gravitational release is $\sim$10\% (see Shakura \& Sunyaev 1973).
This is $\leq$1\%  of our derived mass-loss rate of \#41. 
However,  Table~\ref{ngc300x1} also shows that the WR star would 
completely fill its Roche lobe for the 26  $M_{\odot}$ case ($i 
\leq$35$^{\circ}$ is excluded), and equates to 80\% 
of  its Roche lobe radius, $r_{\rm L}$ (Eggleton 1983), for the 15
$M_{\odot}$ case. Therefore,  the accretion disk may be fed 
primarily through Roche lobe overflow. For comparison, 
the  higher temperature obtained for the WN star in IC\,10 X--1 by Clark
\& Crowther (2004) would favour a wind-fed accretion disk, since the WR 
radius is  $\sim\,0.5 r_{\rm L}$ in that system.

%
%


IC\,10 X--1 is an eclipsing X-ray system (Prestwich et al. 2007), 
therefore geometric arguments 
imply that the black hole would be eclipsed  for $i \geq 78^{\circ}$
for the WR properties derived by Clark \& Crowther (2004). If we adopt a 
radius of $\sim 0.5 r_{\rm L}$ for the accretion  disk, an eclipse of the 
X-ray emitting accretion disk would require $i \geq 80^{\circ}$. NGC\,300 
X--1 does exhibit significant
X-ray variability, but lacks a deep X-ray eclipse (Carpano et al. 2007b). 
Therefore,  geometric  arguments appear to rule out inclinations that
would cause a total eclipse of the accretion disk ($i  \leq 73\pm2 
^{\circ}$). However, a glancing  eclipse  would require  $i \geq 63.5\pm 
3.5^{\circ}$  for the range of  WR  radii obtained here. We therefore 
adopt $i = 80-90^{\circ}$ for IC\,10  X--1 and $i = 60-75^{\circ}$ for 
NGC\,300 X--1.

\begin{figure}
\begin{center}
\includegraphics[width=0.7\columnwidth,angle=270,clip]{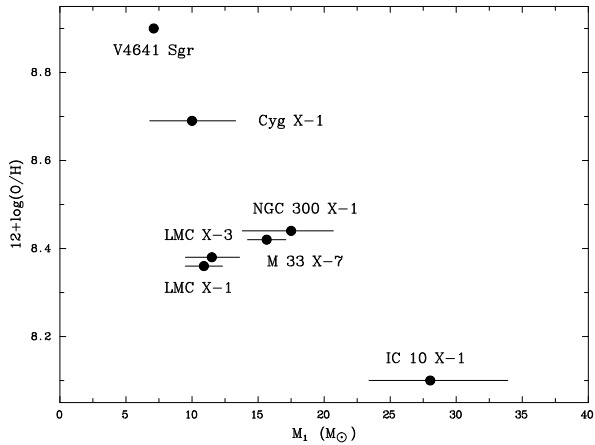}
\caption{Comparison between inferred compact object masses, $M_{1}$,
versus metallicity for all HMXB with $M_{1} \geq 3 M_{\odot}$
and $M_{2} \geq 5 M_{\odot}$. Black hole masses inferred for 
NGC\,300 X--1 (IC\,10 X--1) relate to a WR mass of 21$^{+5}_{-6}$ 
$M_{\odot}$  (25$^{+13}_{-8}$ $M_{\odot}$) and an orbital  inclination of 
60--75$^{\circ}$  (80--90$^{\circ}$).}
\label{mass}
\end{center}
\end{figure}

In Fig.~\ref{mass} we present the host galaxy metallicity as a function of
black hole masses for NGC\,300 
X--1 and IC\,10 X--1, plus those for all HMXB systems for which the 
presence of  a black hole is unambiguous, whose companion is
an OB star with mass $M_{2} \geq5 M_{\odot}$, i.e. LMC X--3 
(Val-Baker et  al. 2007),  LMC X--1  (Orosz, Steeghs \& McClintock 
2009), M33 X--7 (Orosz  et al.  2007), V4641 Sgr (Orosz et al. 2001) 
and Cyg X--1 (Gies 
et al. 2003). We limit our sample to  classical HMXB, to ensure that their 
present-day (oxygen) metallicities 
are  consistent with  the formation of these black hole binaries. The 
majority  of black holes in Low Mass X-ray Binary (LMXB) systems have 
masses close to $10 M_{\odot}$ (Remillard \& McClintock 2006).

It may be significant that both Wolf-Rayet/black hole systems are located 
in metal-poor galaxies. IC~10 has an oxygen content of $\log$ (O/H) + 12 = 
8.1 (Garnett 1990) while NGC\,300 X--1/\#41 is located at a de-projected 
distance of 0.43 $\rho_0$, where $\rho_0$ = 9.75 arcmin (Schild et al. 
2003). According to Urbaneja et al. (2005), the oxygen content at this 
galactocentric distance in NGC\,300 is $\log$ (O/H) + 12 $\sim$ 8.44, 
i.e. relatively similar to the LMC for which $\log$ (O/H) + 12 $\sim$ 8.37 
(Russell \& Dopita 1990). The only other HMXB whose
black hole mass is known to greatly exceed 10 $M_{\odot}$ is M 33 X-7
(Orosz et al. 2007), for which a near identical oxygen content of 
$\log$ (O/H) + 12 = 8.42 is inferred at its location in M 33 
from the calibration of Magrini et al. (2007).



High black hole masses require that the progenitor star was very massive 
and experienced low mass-loss rates (Belczynski et al. 2009). Weak stellar 
winds is a natural consequence of low metallicity 
(Mokiem et al. 2007). However, orbital periods of IC\,10 X--1 
and NGC\,300 X--1 are so short that the radius of the black hole 
progenitor star must have been larger than the present separation of the 
components. As such, the progenitor would have experienced extreme 
mass-loss through Roche-lobe overflow. Therefore, reconciling high black 
hole masses with close orbital separations is a major challenge for binary 
evolution models. 

In the standard picture, such systems involve a 
common-envelope phase, which would naturally lead to a merger 
(Podsiadlowski, Rappaport \& Han 2003). Alternatively, de Mink et al. 
(2009) propose that the short orbital period results in tidal-locking of 
the stellar rotation, causing a chemically homogeneous evolution through 
rotational mixing (Maeder 1987). In this scenario, binary components would 
remain compact and so circumvent the high mass transfer rates of 
Roche-lobe overflow systems.

If NGC\,300 X--1 and IC~10 X--1 were to survive their second supernova 
explosion, they would form binary black hole systems, merging on a 
timescale of a few Gyr. Binary black hole mergers has been considered by 
Sadowski et al. (2008), who argued that their detection rate may be much 
higher than double neutron star systems for current 
gravitational wave experiments.



In conclusion, new VLT/FORS2 time-series spectroscopy of the WN star \#41 
in NGC\,300 is presented, which confirm that it is physically associated 
with the NGC\,300 X--1 system. 
%
We find that NGC\,300 X--1 hosts the most massive stellar-mass 
black hole known, with the exception of the other extragalactic WR/black 
hole system IC~10 X--1.

\section*{Acknowledgements}
We wish to thank John Hillier for maintaining CMFGEN, 
and the referee for suggesting improvements to the original
manuscript.



\label{lastpage}
\end{document}